\documentclass[prl,twocolumn,preprintnumbers,footinbib,tightenlines,superscriptaddress]{revtex4}

\pdfoutput=1
\usepackage[english]{babel}
\usepackage{amsmath,amssymb,amsfonts,bm,bbm,slashed,subdepth}
\usepackage{graphicx}
\usepackage[sort&compress]{natbib}
\usepackage{xcolor}
\usepackage[normalem]{ulem}
\usepackage{hyperref}
\usepackage{cleveref}
\definecolor{red}{rgb}{1.0, 0, 0}
\usepackage{enumerate}
\usepackage{epsfig, subfigure}
\usepackage{setspace}
\usepackage{booktabs, tabularx}
\usepackage{units}
\usepackage{placeins}
\usepackage{multirow}
\usepackage{mathtools}

\newcommand{\DM}{\text{DM}}

\begin{document}

\preprint{DESY 18-225}
\preprint{IPMU18-0207}
\title{Finite-size dark matter and its effect on small-scale structure}

\author{Xiaoyong Chu}
\email{xiaoyong.chu@oeaw.ac.at }
\affiliation{Institute of High Energy Physics, Austrian Academy of Sciences, Nikolsdorfer Gasse 18, 1050 Vienna, Austria
}

\author{Camilo Garcia-Cely}
\email{camilo.garcia.cely@desy.de}
\affiliation{Deutsches Elektronen-Synchrotron DESY, Notkestrasse 85,
22607 Hamburg, Germany}
\author{Hitoshi Murayama}
\email{hitoshi@berkeley.edu, hitoshi.murayama@ipmu.jp}
\affiliation{Department of Physics, University of California, Berkeley, CA 94720, USA}
\affiliation{Kavli Institute for the Physics and Mathematics of the
  Universe (WPI), University of Tokyo,
  Kashiwa 277-8583, Japan}
\affiliation{Ernest Orlando Lawrence Berkeley National Laboratory, Berkeley, CA 94720, USA}
\affiliation{Deutsches Elektronen-Synchrotron DESY, Notkestrasse 85,
22607 Hamburg, Germany}

\begin{abstract} 
If dark matter has a finite size that is larger than its Compton wavelength, the corresponding self-interaction cross section decreases with the velocity. We investigate the implications of this Puffy Dark Matter  for addressing the small-scale problems of the $\Lambda$CDM model, and  show that the way the non-relativistic cross section varies with the velocity is largely independent of the dark matter internal structure. Even in the presence of a light particle mediating self-interactions, we find that the finite-size effect may dominate the velocity dependence.
  We present an explicit example in the context of a QCD-like theory  and discuss possible ways to differentiate Puffy Dark Matter from the usual light-mediator scenarios. 
  Particularly relevant for this are low-threshold direct detection experiments and indirect  signatures associated with the internal structure of dark matter. 
 \end{abstract}

\maketitle

{\bf Introduction.} The dark matter (DM) nature is one of the most important open questions of our century.  Until now, we have only observed DM via its gravitational effects, with  data supporting the hypothesis  that DM is collisionless at large  scales~\cite{Aghanim:2018eyx}. This is at the core of the celebrated $\Lambda$CDM model and its most stringent tests come from large objects such as clusters of galaxies, which constrain the self-scattering cross section per unit mass, $\sigma/m$, to be below $\unit[1]{cm^2/g}$~\cite{Randall:2007ph,Robertson:2016xjh}. Despite this, larger values of $\sigma/m$ are not ruled out in small objects  such as galaxies and dwarf spheroidals. Nonetheless, that requires the self-scattering cross section to decrease with the DM velocity,  because the particles residing in larger DM halos move faster.  
This paradigm is known as self-interacting dark matter (SIDM)~\cite{Spergel:1999mh} and has attracted a lot of attention from astronomers and particle physicists in the last two decades. 

One reason for this is the apparent  mass deficit in the inner regions of small-scale halos with respect to the predictions of collisionless DM. This has led to the so-called small-scale crisis of the $\Lambda$CDM model, which might be solved by SIDM because it predicts DM halos with smaller central densities~\cite{Vogelsberger:2012ku,Rocha:2012jg,Peter:2012jh}. For a review  see~\cite{Tulin:2017ara, Bullock:2017xww}. Another reason for the continued interest in SIDM is that it gives clues about specific properties of DM, which can be used to search for it. For instance, large and velocity-dependent cross sections  might hint at a long-range force, which in turn suggests the presence of a light mediator. In fact, since such a particle is a rather generic feature of several well-motivated DM models,  velocity-dependent SIDM is often associated with a light mediator.  In this work we discuss another source of velocity dependence for $\sigma/m$ (see also ~\cite{Feng:2009hw,Tulin:2013teo,Chu:2018fzy,McDermott:2017vyk, Vogelsberger:2018bok,  Chu:2018nki}), which hints  at DM particles of finite size, $r_\DM$.

\begin{figure}
\includegraphics[width=0.47\textwidth]{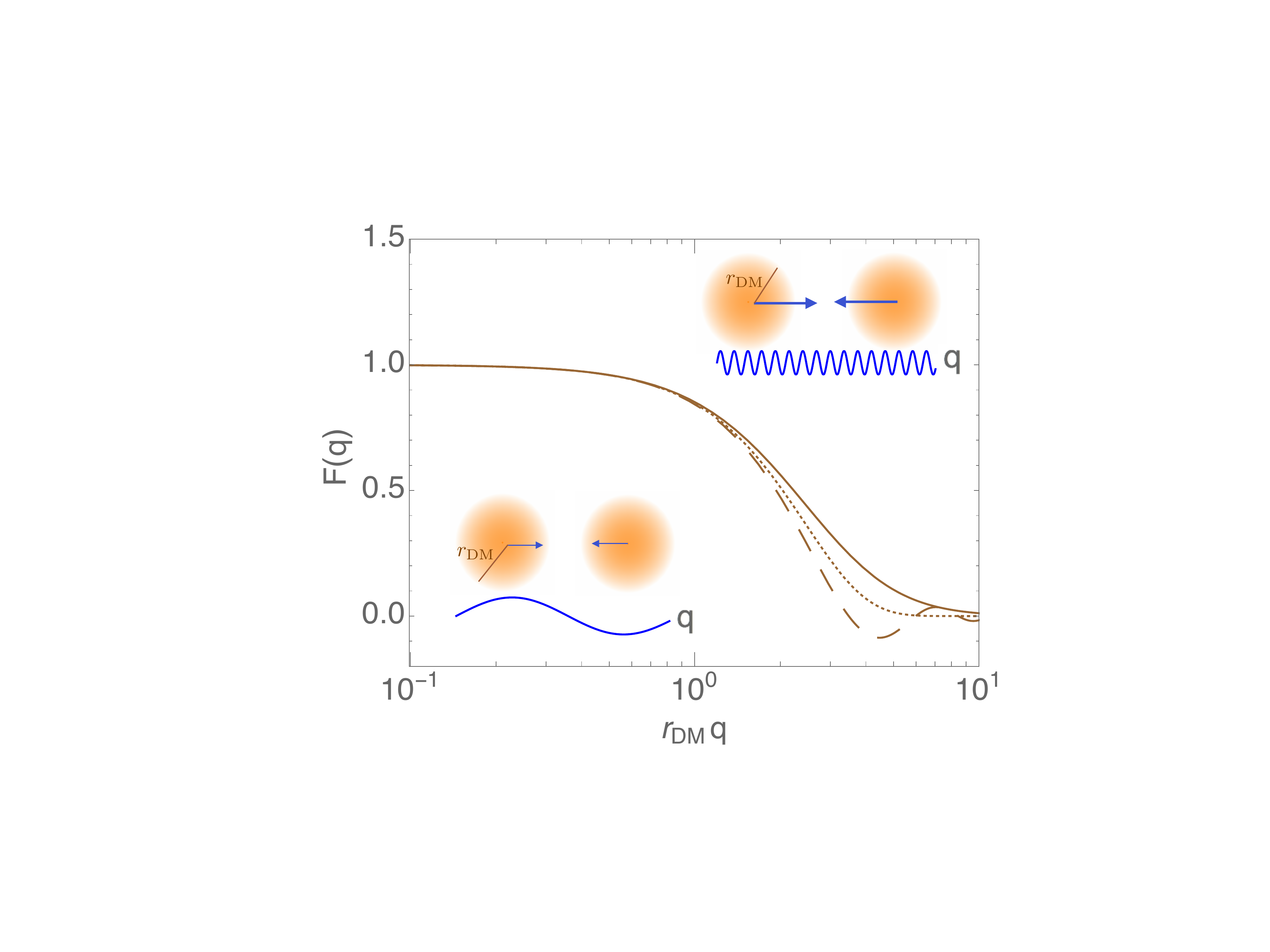}

\caption{Form factors as a function of momentum-transfer $q$ in units of the inverse root-mean-square radius $r_\text{DM}$. Solid, dashed and dotted lines correspond to the dipole, tophat and Gaussian distributions (see Table~\ref{table:Fq}).
}
\label{fig:fi2}
\end{figure}

As is shown in Fig.~\ref{fig:fi2}, a momentum transfer much smaller than $r_\text{DM}^{-1}$  is too small to measure the internal structure of the DM, so the latter acts as a point-like particle. 
   On the other hand, 
  when the momentum transfer becomes larger than $r_\text{DM}^{-1}$, the internal structure of the particle is probed. 
As specified below, 
this can happen  in such a way that the phase difference among the scattered waves   
 leads to a suppression in $\sigma/m$. 
 This is indeed the desired velocity dependence. 
In fact, as we will see, even in the presence of light mediators, the finite size may be the dominant effect. 

We will refer to this scenario as Puffy DM. 
Beside the self-scattering effects, the fact that DM has a finite size leads to a very rich phenomenology, as has been explored for several concrete DM candidates~(e.g.\,\cite{Coskuner:2018are, Nussinov:1985xr,Chivukula:1989qb,Kaplan:2009de,	Feldstein:2009tr, Kumar:2011iy, Laha:2013gva, Cline:2013zca, Wise:2014jva, Krnjaic:2014xza, Wise:2014ola, Detmold:2014qqa, Hardy:2014mqa, Hardy:2015boa, Laha:2015yoa, Mitridate:2017oky, Gresham:2018anj, Francis:2018xjd, Contino:2018crt, Braaten:2018xuw, Ibe:2018juk,Ibe:2018tex}).

{\bf II. Scattering of finite-size DM particles.} Let us first consider the scattering of two finite-size objects, which --for simplicity-- will be  modeled as a collection of point-like constituents that coherently scatter  by means of a spin-independent Yukawa interaction. The corresponding charge density, $\rho(\vec{\bf r})$, characterizes the finite shape of the scattering object. We will also assume that the contribution of the binding force to the scattering rate  is  negligible. This is the case e.g. if such a force leads to a momentum-suppressed scattering amplitude.
 Then,   
the interaction Hamiltonian for  two objects described by the density profiles $\rho_{1}(\vec{\bf x})$ and $\rho_{2}(\vec{\bf y})$ is
\begin{eqnarray}
	H_{\it int} &=& 
	\int d \vec{\bf x} d \vec{\bf y} \rho_{1}(\vec{\bf x}) 
	\frac{ \alpha e^{-|\vec{\bf x}-\vec{\bf y}|/\lambda }}{ |\vec{\bf x}-\vec{\bf y}|} \rho_{2}(\vec{\bf y})\nonumber\\
&=& \int \frac{d \vec{\bf q}}{(2\pi)^{3}} F_{1} (\vec{\bf q}) 
	\frac{ 4\pi \alpha}{\vec{\bf q}^{2}+\lambda^{-2}} F_{2} (-\vec{\bf q})\,.
\label{eq:Yuk}
\end{eqnarray}
where $\lambda$ is the range of the interaction, $\alpha$ is a coupling constant, and we have introduced  the form factor $F_i(\vec{\bf q}) \equiv \int  {d \vec { \bf r} \, e^{i \vec {\bf  q} \cdot \vec { \bf r}} } \rho_i(\vec{\bf r})$.  
Hence, the center-of-mass differential cross section in the Born approximation is
\begin{equation}
	\frac{d\sigma}{d\Omega} = S \left|
	F_{1} (\vec{\bf q}) \frac{2  \mu  \alpha}{\vec{\bf q}^{2} +\lambda^{-2}} F_{2} (-\vec{\bf q}) \,\pm\, (\vec{\bf q}\to-\vec{\bf q}) \right|^{2},
\label{eq:scattering1}
\end{equation}
where $\mu$ is the reduced mass and $\vec{\bf q}$ is the momentum transfer. For identical (non-identical) particles, the second term must (not) be included and $S=1/2\, (1)$. 

\begin{table}
\begin{tabular}{cccc}\hline
Shape &  $\rho(r)$& $r_\text{DM}$ & $F(q)$\\\hline
tophat &$ \frac{3}{4\pi r_0^3} \theta(r_0-r)$ & $2\sqrt{3}r_0$ & $\frac{3 (\sin (r_0 q)-r_0 q \cos (r_0 q))}{r_0^3 q^3}$  \\
dipole &  $\frac{e^{-r/r_0}}{8\pi r_0^3}  $  & $\sqrt{3/5}r_0$ & $\frac{1}{\left(1+r_0^2 q^2\right)^{2}}$ \\
Gaussian  & $ \, \frac{1}{8 r_0^3 \pi^{3/2}} e^{-r^2/(4r_0^2) }$ & $\sqrt{6} r_0$ & $e^{-r_0^2q^2}$\\\hline
\end{tabular}
\caption{Form factors for different density distributions. }
\label{table:Fq}
\end{table}

An illustrative example is the electron scattering off finite-size objects. This is determined by a Coulomb interaction ($\lambda \to \infty$) with $\rho_e (\vec{\bf r}) = \delta(\vec{\bf r})$ or $F_e(\vec{\bf q}) =1$. In this case, Eq.~\eqref{eq:scattering1}  gives the well-known Rutherford scattering formula, which can be used to infer the shape of finite-size objects.  When applied to the proton, one finds   a density distribution decreasing exponentially with a characteristic scale $r_0^{-2} = 0.71$\,GeV$^2$~\cite{Perdrisat:2006hj}.  The latter is the dipole distribution  (see Table~\ref{table:Fq}),  generally expected from wave-function solutions to various potential wells~\cite{Landau:1991wop}.

\begin{figure}
\includegraphics[width=0.45\textwidth]{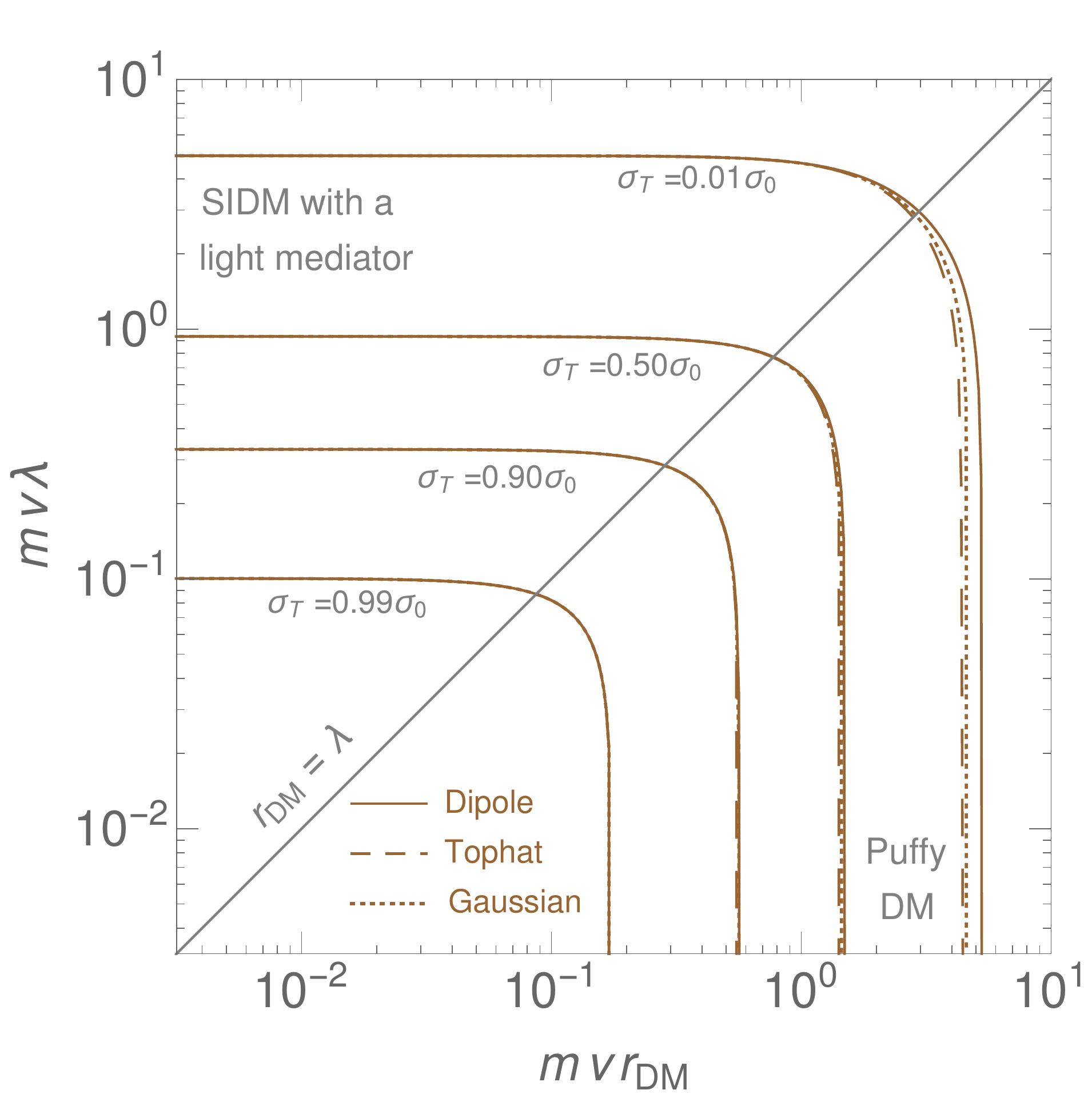}
\caption{Transfer cross section as a function of the force range, $\lambda$, and the DM size, $r_\text{DM}$, both in units of $1/mv$. Here $\sigma_0$ of Eq.~\eqref{eq:ansatz} is assumed to be constant. 
}
\label{fig:fourier}
\end{figure}

We apply now Eq.~\eqref{eq:scattering1} to non-relativistic DM.  Assuming that the DM particle is spherical, i.e. $F(\vec{\bf q})=F(q)$, the $S$-wave differential cross section reads
\begin{equation}\label{eq:ansatz}
	\frac{d \sigma}{d \Omega}= \frac{\sigma_{0} }{8\pi}
	\left[ \frac{F(q)^2}{1+\lambda^2q^2} + \left( \theta \to \pi-\theta \right)\right]_{q= m v\sin\theta/2}^2 \,,
\end{equation}
where $\sigma_0= 4 \pi (m \alpha \lambda^2)^2 $. Here $\theta$ and $v$ are respectively the scattering angle and the relative velocity in the center-of-mass frame.  While the exact form of $\rho(r)$ --and hence $F(q)$ in Eq.~\eqref{eq:ansatz}-- needs to be determined by solving for the wave function from the Schr{\"o}dinger equation of
 the composite state,  the differential cross section is
not sensitive to the details of $\rho(r)$ as long as it is always positive (no screening) and it goes to zero sufficiently fast at large radii. In that case, the DM size --or more precisely-- the root-mean-square radius 
\begin{equation} 
r^2_\text{DM}\equiv \int d \vec{\bf r}\,  \rho(r) r^2 =-6 \,\frac{d^2 F(q)}{dq^2} \Bigg|_{q=0}  \,
\label{eq:size}
\end{equation}
is positive. Thus, $F(q)$ decreases for small momenta from $F(0)=\int {d \vec{\bf r}} \rho (r)$, which can be normalized to $1$ without loss of generality.
Fig.~\ref{fig:fi2} illustrates this for the three representative distributions as listed in Table~\ref{table:Fq}.  
 Together with Eq.~\ref{eq:ansatz}, all this implies  that the cross section is constant at low velocities and eventually approaches zero, even if the range of the interaction is short.

{\bf III.  DM scattering in astrophysical halos.} 
Because of the form factor, for low velocities we expect isotropic scattering, whereas for larger velocities forward scattering is more probable. Due to this,
the transfer cross section, $\sigma_T \equiv \int d\Omega (1-|\cos\theta|)d\sigma/d\Omega$, captures the self-interaction effects in DM halos  better  than $\sigma$ (see e.g. \cite{Kummer:2017bhr}), and will be adopted below.  

Fig.~\ref{fig:fourier} illustrates the dependence of $\sigma_T$ on the interaction range $\lambda$ and the particle size $r_\text{DM}$. As apparent from the plot, $\sigma_T$ is largely independent of the exact expression for the form factor and therefore of $\rho(r)$. Furthermore,  roughly speaking, the transfer cross section is constant for $ mv\ll \text{min}\{\lambda^{-1}, r_\text{DM}^{-1}\}$, starts decreasing at $ mv \sim \text{min}\{\lambda^{-1}, r_\text{DM}^{-1}\} $, and approximately scales as $1/v^4$  for $mv\gg \text{min}\{\lambda^{-1}, r_\text{DM}^{-1}\}$. 
This directly follows from rewriting the transfer cross section as 
\begin{equation}
\sigma_T   
	= \hspace{-5pt}\int^{\frac{(mv)^2}{2}}_0   \hspace{-3pt} dq^2
	\left[  \dfrac{ \,F(q)^2}{1+\lambda^2q^2} + \hspace{-3pt}\left(q^2 \to (mv)^2-q^2  \right) \right]^2\hspace{-3pt}\frac{2\sigma_0q^2 }{(mv)^4}\,.
\end{equation}

See the Appendix for details.
When the range of the Yukawa force is much larger than the DM size, such a velocity dependence of $\sigma_T$ coincides with that of the Born regime of SIDM with a light mediator~\cite{Tulin:2013teo}. In fact, Fig.~\ref{fig:fourier} shows that there is a one-to-one correspondence between the latter and the self-scattering of finite-size DM. Furthermore,  there could be a mediator  lighter than the DM and still the velocity dependence is determined by the DM size if $\lambda\lesssim r_\text{DM}$. 

\begin{figure}[t]
\includegraphics[width=0.49\textwidth]{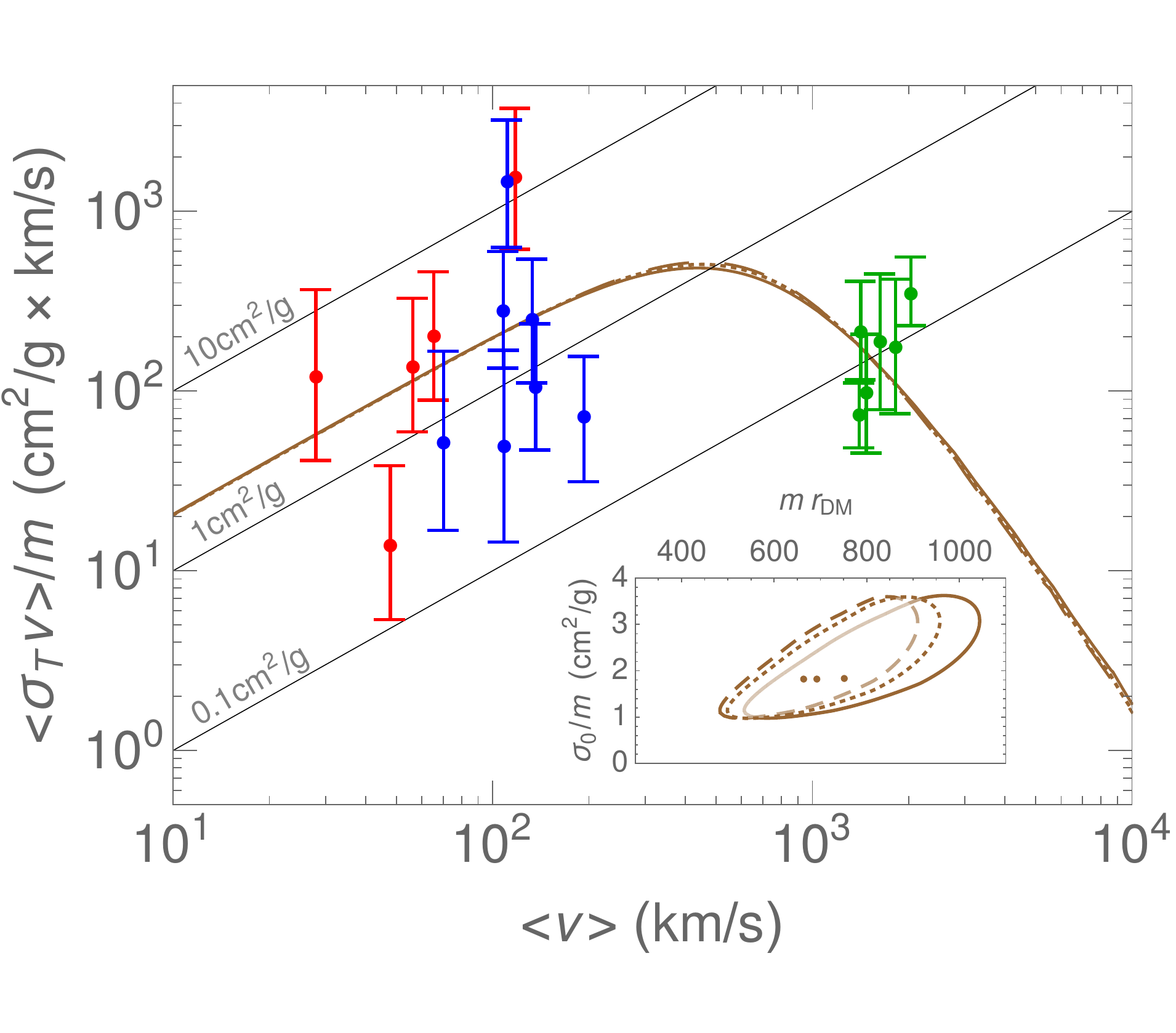}
\caption{Velocity dependence of the transfer cross section of Puffy DM. Best-fit curves to data \cite{Kaplinghat:2015aga} for the dipole (solid), tophat (dashed) and the Gaussian (dotted) distributions in Table~\ref{table:Fq}. The inset shows the 95\% C.L. contours on the parameter $\sigma_0$ from Eq.~\eqref{eq:ansatz} and the DM size   together with the corresponding parameter sets plotted in the main  figure.
}
\label{fig:fit}
\end{figure}

The  DM relative velocity in astrophysical halos typically follows  a Maxwell-Boltzmann distribution truncated at the corresponding escape velocity,  $v_\text{max}$.  The velocity-averaged transfer cross section is then~\footnote{Here we will assume that $v_\text{max} \to \infty$ since the integral converges quite fast due to  the Boltzmann  factor. 
}
\begin{equation}
\langle \sigma_T v \rangle = \int^{v_\text{max}}_0 f(v) \sigma_T v dv \,,\quad f (v ) = \frac{32 v^2e^{-{4v^2}/{\pi \langle v \rangle ^2}}}{ \pi^2 \langle v \rangle^3} \,.
\label{eq:sigmaTv}
\end{equation}
In the context of SIDM as a solution to the small-scale structure problems, a semi-analytical method has been proposed in \cite{Kaplinghat:2015aga} to infer,  from  observational data, the   value of $\langle \sigma_T v\rangle/m$ for a given DM halo (see also \cite{Valli:2017ktb}). This  method was applied to five clusters from \cite{Newman:2012nw},  seven low-surface-brightness spiral galaxies  in \cite{KuziodeNaray:2007qi} and six  dwarf galaxies of the HI Nearby Galaxy Survey  sample~\cite{Oh:2010ea}. Fig.~\ref{fig:fit} shows these results respectively in green,  blue and red. The  set of points    
is also in agreement with  cluster bounds mentioned in the Introduction, giving $\sigma_T/m\lesssim\unit[1.3]{cm^2/g}$~\cite{Randall:2007ph,Robertson:2016xjh}. While cosmological simulations show this semi-analytical method works for isolated halos (see e.g.~\cite{Vogelsberger:2015gpr, Creasey:2016jaq, Sokolenko:2018noz}), recent studies suggest that tidal stripping may further modify the density profile of satellite halos~\cite{Kaplinghat:2019svz, Sameie:2019zfo, Zavala:2019sjk, Kahlhoefer:2019oyt}. Such effects are not included here, because the galaxies shown in Fig.~\ref{fig:fit} are in the field.

Postulating a DM finite size much larger than the range of the Yukawa force, i.e. $\lambda\ll r_\text{DM} $, provides an excellent fit to the velocity-dependent cross section preferred by the galactic and cluster systems.  
The corresponding best-fit of Eq.~\eqref{eq:sigmaTv} to the  data above is shown in Fig.~\ref{fig:fit} for the dipole, tophat and Gaussian distributions, separately. As expected from the aforementioned remarks, there is almost no dependence on details of the form factors even though they correspond to  substantially different density distributions.  The figure also shows that, in order to have the right velocity dependence, the DM size needs to be hundreds of times larger than the Compton wavelength. This explains the name  Puffy DM. 

If the Yukawa force is  associated with a mediator $\rho$, requiring that the finite size dominates the scattering, i.e. $\lambda \lesssim r_\text{DM}$ implies $ m_\rho \gtrsim 10^{-3} m$. This shows that the mediator can still be substantially lighter than Puffy DM. Moreover,   if we impose $\alpha \lesssim m_\rho/m$ as required in the Born expansion,  $\sigma_0/m= 4 \pi (m \alpha \lambda^2)^2/m \sim \unit[1]{cm^2/g} $ leads to $m\lesssim \unit[20]{GeV}$. Consequently, Puffy DM must lie at the GeV scale or below.  

{\bf IV. A model of Puffy DM.}  Here we only sketch a possible realization of Puffy DM while details will be discussed elsewhere.  It is a QCD-like confining  theory with $N_c$ colors and two flavors of quarks: one ``charm quark'' much heavier than the confining scale $\Lambda$ and one nearly massless ``down quark''. They respectively have charges $+2/3$ and $-1/3$ under a dark $U(1)_D$ gauge group with $N_c=3$. This is  associated with a massive ``dark photon'' $\gamma_{D}$, which can act as a portal to the Standard Model (SM) by means of the kinetic mixing between the $U(1)_D$ group and the SM hypercharge. There are no dark weak interactions. We assume there is an asymmetry so that anti-charm quarks are annihilated while the remaining charm quarks end up in the baryonic $\Sigma_{c} (cdd)$ state.  The latter interacts by exchanging the  pseudo-scalar $\eta (d\bar{d})$ and the vector $\rho(d\bar{d})$, which lead to attractive and repulsive forces respectively. 

On the one hand, it is likely that the $\eta$-exchange dominates binding $\Sigma_c$ baryons into nuclei because its range is larger given that the $\eta$ mass is due to the anomaly and hence suppressed as $m_{\eta} \sim \Lambda / \sqrt{N_{c}}$, as opposed to the $\rho$ mesons for which  $m_{\rho} \sim \Lambda$.  In view of this, in the  following we assume the typical mass number is $10 \lesssim A \lesssim 100$.
On the other hand, the nucleus-nucleus scattering is dominated by the exchange of $\rho$ mesons  because the latter are  essentially  massive gauge bosons coupled to $d$-number ($A/2$) giving rise to coherent spin-independent scattering, while the $\eta$-exchange induces a spin-dependent momentum-suppressed scattering.  Therefore, the range of the scattering force $\Lambda^{-1}$ is shorter than the size of the nuclei $r_\text{DM}\sim A^{1/3} m_{\eta}^{-1} \sim A^{1/3} \Lambda^{-1} \sqrt{N_{c}}$. 
As a result, this model is a realization of Puffy DM. 

 For instance, parameters such as $N_c=3$, $A\sim10$, $m_c\sim m_{\Sigma_{c}} \sim \unit[1]{GeV}$, $r_\text{DM}^{-1}\sim \unit[15]{MeV}$, $m_\eta\sim \unit[20]{MeV}$, $\Lambda \sim m_\rho\sim\unit[30]{MeV}$ and $\alpha \sim m_\rho/m$ realize the desired self-scattering cross section and its velocity dependence.  We  take ${\gamma_D}$ to be slightly lighter than  $\eta$. Then the size of the kinetic mixing is either (A) $10^{-5} \lesssim \epsilon \lesssim 10^{-3}$ or (B) $\epsilon \ll 10^{-10}$ to satisfy beam-dump experimental data and supernova  observations~\cite{Hardy:2016kme, Chang:2016ntp, Ibe:2018juk}. In the cosmological history, presumably much of the entropy in this sector ends up in a thermally populated gas of $\eta$ mesons. These decay via  $\eta \rightarrow \gamma_{D} \gamma_{D}^* \rightarrow 2(e^{+} e^{-})$ or self-annihilate via  $\eta\,\eta\to \gamma_{D} \gamma_{D} $ before BBN for the range (A)~\footnote{Assuming that $m_{\gamma_D}\le m_{\eta}$ is not essential. If a dark axion, $a$,  exists, the annihilation $\eta \eta \to \eta a$ can play the same role.}.

\begin{figure}[t]
\includegraphics[width=0.49\textwidth]{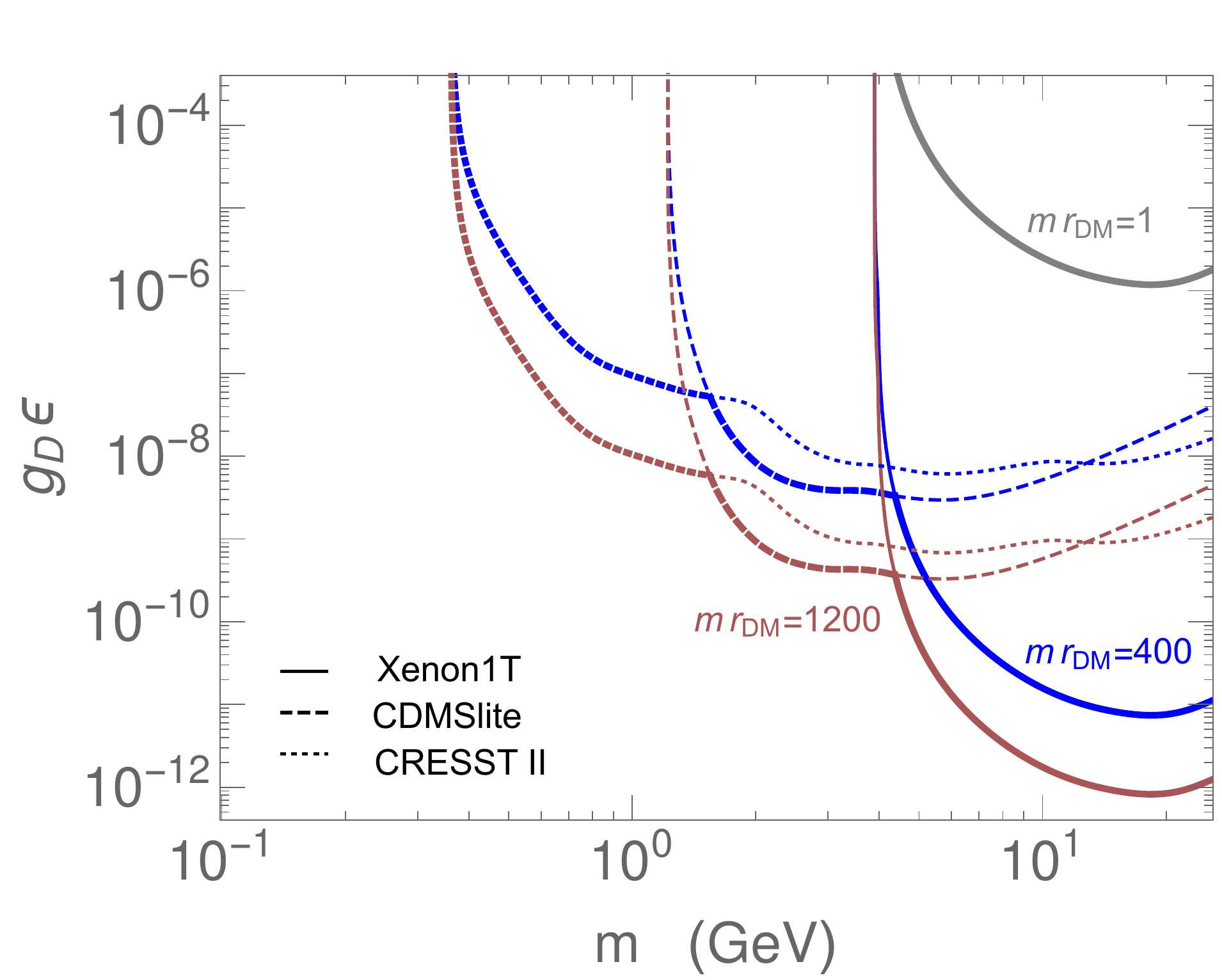}
\caption{Direct detection bounds on our QCD-like theory of Puffy DM from nucleon recoil events in Xenon1T~\cite{Aprile:2017iyp}, CMDSlite~\cite{Agnese:2015nto} and CRESST II~\cite{Angloher:2015ewa}, assuming $m_{\gamma_D}=20$\,MeV. For a heavier dark photon, this bound scales with $m_{\gamma_D}^2$.   
See text for details. 
}
\label{fig:DDbound}
\end{figure}

Direct detection signatures are closely related to the DM finite size.  
The recoil-energy spectrum  is that of a point-like DM particle scattering via a contact interaction
times the square of the dimensionless factor $\xi(q) = F_D(q) \,m_{\gamma_D}^2/(q^2+m_{\gamma_D}^2)$. Its first part is the form factor  associated with the  $U(1)_D$ charge and the second one parametrizes the dependence on the mediator mass. $ F_D(0)=0$  because the DM particle is neutral under $U(1)_D$~\footnote{This is different from self-interaction form factor, where $F(0)=1$, while the charge radii of both should be of the similar size.}. Thus, to leading order in $q$, $\xi(q) \simeq q^2\, F''_D(q=0)/2 $,  which induces  DM scattering rates  enhanced by the fourth power of charge radius. The latter is given by Eq.~\eqref{eq:size} with $F\to F_D$ and is expected to be similar to $r^4_\text{DM}$ on dimensional grounds.  In the view of this, we estimate current direct-detection  limits by implementing such a recoil spectrum in DDCalc~\cite{Athron:2018hpc, Workgroup:2017lvb}.  The results are shown in Fig.~\ref{fig:DDbound} for various choices of the charge radius and  $ m_{\gamma_D}\geq \unit[20]{MeV} $.   For a lighter $\gamma_D$, the bound becomes more stringent than that due to an enhanced $\xi(q)$. Hence,  Xenon1T significantly constrains DM masses above 5 GeV, while low-threshold direct detection experiments such as CRESST-II and CDMSlite can  probe smaller masses.

 A salient aspect of  this DM setup is that energy recoils are momentum-suppressed albeit the enhancement due to the large charge radius. This is in sharp contrast to the direct detection of point-like SIDM by means of  light mediators. Since $q$ is proportional to the reduced mass of the colliding particles,  we expect signals in experiments searching for nuclear recoils but not in those looking for electron recoils, 
whose $\xi(q)$ is much smaller for the DM masses of interest here.

Likewise, the internal structure of Puffy DM  allows for up-scattering processes, giving rise to a wealth of indirect search signatures if DM de-excites ejecting SM particles.  In our QCD-like model this  happens due to kinetic mixing, and for the parameters considered above, the required exciting energy can be estimated as $ O(\Lambda^{2}/m_c)\sim$ MeV. While this is much greater than the typical galactic DM kinetic energy, 
DM might be excited by inelastic self-scatterings  in massive clusters of galaxies  or  by collisions with  high-energy cosmic rays~\cite{DEramo:2016gqz, Cappiello:2018hsu, Bringmann:2018cvk, Cappiello:2019qsw}. The former case may lead to radio and X-ray signals~(e.g. \cite{Storm:2012ty, Storm:2016bfw, Marchegiani:2018xck}) or DM dissipative cooling~(e.g.~\cite{Fan:2013yva,Foot:2013vna,Fan:2013bea, Boddy:2016bbu, Essig:2018pzq}); while the latter might trigger novel signals in direct-detection and neutrino experiments as in e.g.~\cite{Giudice:2017zke, Ema:2018bih}.  A detailed investigation  is beyond the scope of this letter.

Before concluding, we would like to emphasize that Puffy DM does not necessarily require QCD-like dynamics. Indeed, Puffy DM can be realized in other theories of extended objects. For instance, (non-)topological defects, such as Q-balls~\cite{Kusenko:1997si,Kusenko:1997vp,Kusenko:2001vu} or skyrmions~\cite{Mielke:2002bp,Gillioz:2010mr,Kitano:2016ooc},  are naturally stable, have a large size and self-interact. 
The study of Puffy DM in the form of defects is  an ongoing  project. 

{\bf V. Conclusions.} We have shown that if DM is an extended object with a size hundreds of times larger than its Compton wavelength, the corresponding self-interaction cross section  varies with velocity in a way that is largely independent of its internal structure. For cross sections larger than  $\unit[1]{cm^2/g}$  at $v\to 0$, this provides a solution to the problems of the $\Lambda$CDM model in small-scale astrophysical objects while still being in agreement with cluster observations.  A QCD-like theory where DM is a dark nucleon has been used to illustrate our results, which are nevertheless general and can be applied to a broader range of theories. For this reason, we believe Puffy DM opens up a new avenue for SIDM model-building.    

\newpage
\begin{acknowledgements}
{\bf Acknowledgments.} We thank  Fady Bishara, Bob Cahn, Yohei Ema, Ranjan Laha, Kai Schmidt-Hoberg and Sebastian Wild for enlightening  discussions. 
X.C. is supported by the `New Frontiers' program of the Austrian Academy of Sciences. C.G.C. is supported by the ERC Starting Grant NewAve (638528).
H.M. thanks the Alexander von Humboldt Foundation for support while this work was completed.  H.M. was supported by the NSF grant PHY-1638509, by the U.S. DOE Contract DE-AC02-05CH11231, by the JSPS Grant-in-Aid for Scientific Research (C) (17K05409), MEXT Grant-in-Aid for Scientific Research on Innovative Areas (15H05887, 15K21733), and by WPI, MEXT, Japan.  
\end{acknowledgements}

\appendix

\section{Appendix: The transfer cross section}
\label{appendix:sigma}

The transfer cross section for DM scattering is

\begin{widetext}

\begin{equation}
\sigma_T = \frac{\sigma_0}{8\pi}\int d\Omega\,(1-|\cos\theta|)
	\left[ \frac{F(q)^2}{1+\lambda^2q^2} + \left( \theta \to \pi-\theta \right)\right]^2 
	= \int^{\frac{(mv)^2}{2}}_0   
	\left[  \frac{ \,F(q)^2}{1+\lambda^2q^2} + \left(q^2 \to (mv)^2-q^2  \right) \right]^2\frac{2\sigma_0q^2dq^2 }{(mv)^4}. \label{app:eqsT}
\end{equation}

\end{widetext}

Here we focus on the Puffy DM, where $r_\text{DM}^{-1} \ll  \lambda^{-1}$. On the one hand, taking the low velocity limit, $m v \ll r_\text{DM}^{-1} $,  the factor in the square bracket approaches $2$ and thus $\sigma_T \to \sigma_0$ at $v\to 0$. On the other hand, for $m v \gg r_\text{DM}^{-1}$,   $F(q)$ is suppressed at $q \gg r_\text{DM}^{-1}$  so that the result of the integral is  
insensitive to its upper limit. The integral is not sensitive to $\lambda$ either,  because for any $q \gtrsim \lambda^{-1}$ there is always  $q \gg r_\text{DM}^{-1}$.  Taking this into account allows us to approximate $\sigma_T$ by 
 \begin{eqnarray}
\sigma_T  \simeq \frac{2 \sigma_{0} }{m^4v^4}\, \int^\infty_0 dq^2 q^2   F(q)^4  \simeq    {\sigma_{0} \over   (c\, {m v } r _\text{DM})^4 } \,,
\end{eqnarray}
with $c= 0.23,\,3.9,\,0.97 $ for  the tophat, the dipole,  and the Gaussian distributions, respectively. 
Therefore, $\sigma_T $ scales as $1/v^4$ at $m v \gg r_\text{DM}^{-1}$. The behavior derived here is different from that of effective range theories~\cite{Bethe:1949yr,Chu:2019awd}, since the latter applies to each partial wave of the scattering cross section, while our result applies to the total transfer cross section. Note that at very large $mv$ incoherent scattering starts  playing a role. Nevertheless, its contribution is much smaller than $\sigma_T$, and is therefore neglected for simplicity. 
\bibliographystyle{utcaps_mod}

\bibliographystyle{utcaps_mod}

\bibliography{ref}

\end{document}